# Giant Planets at Small Orbital Distances

T. Guillot[1], A. Burrows[2], W.B. Hubbard[1], J.I. Lunine[1], and D. Saumon[1,3]

[1]Department of Planetary Sciences, University of Arizona, Tucson, AZ 85721 (guillot, hubbard, jlunine, dsaumon@lpl.arizona.edu).

[2]Departments of Physics and Astronomy, University of Arizona, Tucson, AZ 85721 (burrows@cobalt.physics.arizona.edu).

[3]Hubble Postdoctoral Fellow.

## ABSTRACT

Using Doppler spectroscopy to detect the reflex motion of the nearby star, 51 Pegasi, Mayor & Queloz (1995) claim to have discovered a giant planet in a 0.05-A.U., 4.23-day orbit. They estimate its mass to be in the range 0.5 $M_J$ to 2 $M_J$, but are not able to determine its nature or origin. Including the effects of the severe stellar insolation implied, we extend the theory of giant planets we have recently developed to encompass those at very small orbital distances. Our calculations can be used to help formulate search strategies for luminous planets in tight orbits around other nearby stars. We calculate the radii and luminosities of such giants for a variety of compositions (H/He, He, $H_2O$, and olivine), evolutionary tracks for solar-composition gas giants, and the geometry of the Hayashi forbidden zone in the gas-giant mass regime. We show that such planets are stable and estimate the magnitude of classical Jeans evaporation and of photodissociation and loss due to EUV radiation. Even over the lifetime of the primary, the companion would not have lost a large fraction of its mass. In addition, we demonstrate that for the mass range quoted, such planets are well within their Roche lobes. We show that the strong composition-dependence of the model radii and distinctive spectral signatures provide clear diagnostics that might reveal 51 Peg B's nature, should interferometric or adaptive-optics techniques ever succeed in photometrically separating planet from star.

## 1. Introduction

As the search for planets and brown dwarfs around nearby stars accelerates, we should expect to be surprised. In no instance is this better illustrated than in the recent discovery by Mayor & Queloz (1995) of a planet orbiting a G2.5 star, 51 Pegasi, 14 parsecs away. With a 4.23–*day* period, a semi-major axis of 0.05 A.U., an eccentricity less than 0.15, and an inferred mass between 0.5 and 2 *Jupiter* masses ($M_J$), this object is surely the most problematic find in recent memory. As of this writing, the tell-tale periodic Doppler shift in the spectral lines of the primary had been confirmed by Marcy & Butler (1995) and by Noyes *et al.* (1995). A good case can be made for the existence of 51 Peg B simply from the absence of significant photometric variations in $V$ ($< 0.002$ *mag*; Mayor & Queloz 1995; Burki, Burnet, & Kuenzli 1995) or asymmetries in the line profiles (Mayor & Queloz 1995) and from the difficulty of explaining such a period as pulsation of a near-solar analog.

One hundred times closer to its primary than Jupiter itself, 51 Peg B thwarts conventional wisdom. Boss (1995) had argued that the nucleation of a H/He-rich Jovian planet around a rock and ice core could be achieved in a protostellar disk only at and beyond the ice point (at ∼160 Kelvin) exterior to 4 A.U. Walker *et al.* (1995) had surveyed for reflex motion 21 G-type stars over 12 years and seen nothing more massive than 1–3 $M_J$ interior to ∼6 A.U. Zuckermann, Forveille, & Kastner (1995) had measured CO emissions from a variety of near–T Tauri disks, had extrapolated to $H_2$, and had concluded that there may not be enough mass or time to form a Jupiter around a majority of stars. The discovery of 51 Peg B, while not strictly inconsistent with any of these papers, vastly enlarges the parameter space within which we must now search.

Several scenarios for the origin of 51 Peg B are emerging:

1. It could be a canonical gas giant that formed many A.U.'s from 51 Peg A, but through frictional and tidal effects spiraled inward during the protostellar phase (Lin & Bodenheimer 1995);

2. As above, it could have formed conventionally, but collided with a massive companion and lost 90% of its orbital angular momentum and 99% of its orbital energy;

3. It could be composed predominantly of hydrogen and helium accreted from the protostellar disk, but have nucleated in situ around a large rock core (without ice).

4. It could be a giant terrestrial planet formed by the accumulation of planetesimals; or

5. It could be an evaporated, ablated, or tidally stripped brown dwarf or star.

However, whatever the provenance or evolutionary history of 51 Peg B, a knowledge of the thermal and structural characteristics of giant planets with a variety of compositions and masses ($M_{\rm p}$) is required to understand it and others like it. A "chondritic", "helium," or "ice" planet with a mass of $\sim$1 $M_{\rm J}$ has a radius ($R_{\rm p}$) that is significantly smaller than that of a hydrogen-rich Jupiter. At a given effective temperature, smaller radii translate into smaller luminosities ($L$).

Recently, we have studied the theoretical evolution of gas giants around nearby stars with masses from 0.3 through 15 $M_{\rm J}$ and of Brown Dwarfs/M Dwarfs with masses from 10 through 250 $M_{\rm J}$ (Saumon *et al.* 1996, Burrows *et al.* 1995, Saumon *et al.* 1994, Burrows *et al.* 1993, Burrows, Hubbard, & Lunine 1989). Over these three orders of magnitude in mass, the basic input physics is the same. Though we had previously considered the effects of stellar insolation, we had not explored such effects at separations near those of 51 Peg B. In this paper, we present a theory of extra-solar giant planets at small orbital distances ($D$). We calculate the radii and luminosities of planets with a variety of compositions, masses, and separations. Though we focus on the 51 Peg A/51 Peg B system, our results can be extended and scaled to planetary systems with other characteristics and are meant to aid in the formulation of search strategies around nearby stars. In addition, we explore the possibilities of tidal truncation and evaporation and conclude with a discussion on the photometric discriminants of the various theories concerning the nature of 51 Peg B.

## 2. The Radii of Giant Planets as a Function of Composition

The hydrogen/helium equation of state that we have employed for this study is described in Saumon, Chabrier, & Van Horn (1995) and incorporates state-of-the-art prescriptions for the interactions among $H_2$, H, protons, and electrons and for the metallization of hydrogen/helium mixtures at high pressures. The evolutionary codes that we have employed are a Henyey code (Burrows *et al.* 1989, 1993) and a code that treats the quasi-static problem as an implicit two-point boundary value problem (Guillot & Morel 1995). The latter was constructed to address the possibility that Jupiter and Saturn may have non-adiabatic structures (Guillot *et al.* 1995) and allows for the existence of large radiative zones. We have expanded this more general code to address the evolution of giant planets very near their primaries for a variety of planet masses. For the evolution of the gas giants, we used the opacities of Alexander & Ferguson (1994).

At the distance of Jupiter from the Sun, the difference between convective and radiative/convective models is slight. However, when the orbital distance of a gas giant from a G2V star (for example) is smaller than $\sim 0.5$ A.U., external heating by the star becomes important *early* in the planet's evolution towards its steady state. Due to the very small heat diffusivities in gas giants at high pressure, this radiative zone does not penetrate deeply in mass (perhaps encompassing 0.1%), but can penetrate deeply in radius (by as much as 10% in a Hubble time). Convective heat transport from the inner convective zone continues to cool it. The entropy of the radiative zone at the photosphere is maintained at a higher, roughly constant, value as the effective temperature ($T_{\rm eff}$) of the planet stabilizes. Since its interior continues to cool and lose its thermal pressure support, the entire planet continues to shrink. Because the planet's effective temperature is stabilizing, its luminosity is decreasing, though at a progressively lower rate (see §3 below). For 51 Peg B, the predicted radii after 1 Gigayear (Gyr) are between 1.35 $R_{\rm J}$ and 1.9 $R_{\rm J}$ for $M_{\rm p}$'s from 2.0 $M_{\rm J}$ to 0.5 $M_{\rm J}$. These are as much as a factor of two smaller that the corresponding radii for fully convective planets. After 8 Gyr (the estimated age of 51 Peg A), the radii for these same planets are between 1.2 $R_{\rm J}$ and 1.4 $R_{\rm J}$. The behavior of radius versus mass for giant planets in the mass range suggested for 51 Peg B is depicted in Figure 1. Radii and bolometric luminosities (not including the reflected component) for various representative models are given in Tables 1a & 1b. We have included on Figure 1 the corresponding curves for helium, $H_2O$, and olivine ($Mg_2SiO_4$) planets, as well as that for fully convective planets. The equations of state for $H_2O$ and olivine (as representative of rock) were taken from the ANEOS compilation (Thompson 1990). The large mean molecular weight of giant rocky planets ensures that thermal effects are small in most of the interior (Hubbard 1984), so that an olivine or ice planet in close orbit around a star will probably not have a radius significantly larger than predicted by our calculations.

Though 51 Peg A has not had enough time to synchronize its spin period with 51 Peg B's orbital period, 51 Peg B's spin period is surely tidally locked with its orbit at 4.23 days. The time for the tidal spin-down of the planet is given by

$$\tau \sim Q\left(\frac{R_{\rm p}^3}{GM_{\rm p}}\right)\omega_{\rm p}\left(\frac{M_{\rm p}}{M_\star}\right)^2\left(\frac{D}{R_{\rm p}}\right)^6, \tag{1}$$

where $Q$ is the planet's tidal dissipation factor, $\omega_p$ is the planet's primordial rotation rate, $M_\star$ is the star's mass, and $G$ is the gravitational constant. Taking $Q \sim 10^5$ and $\omega_p \sim 1.7 \times 10^{-4}$ s$^{-1}$ (Jupiter's values), we obtain $\tau \sim 2 \times 10^6$ years. For a giant terrestrial planet, $R_p$ is smaller, but $Q$ is also smaller, and so $\tau$ would not be very different. Therefore, since 51 Peg A's age is $\sim 10^{10}$ years, 51 Peg B should always present the same face to its primary, whatever its composition.

The equilibrium effective temperature of the planet is given by the formula:

$$T_{eq} \sim T_\star (R_\star/2D)^{1/2}[f(1-A)]^{1/4}, \qquad (2)$$

and the equilibrium luminosity by:

$$L_{eq} \sim L_\star(1-A)(R_p/2D)^2, \qquad (3)$$

where $R_\star$, $T_\star$, and $L_\star$ are the primary's radius, effective temperature, and luminosity and $A$ is the Bond albedo of the planet, which for Jupiter is $\sim$0.35. The reflected luminosity is $L_{eq}A/(1-A)$. The factor, $f$, is 1 if the heat of the primary can be assumed to be evenly distributed over the planet and 2 if only one side reradiates the absorbed heat. For 51 Peg B and an albedo of 0.35, $T_{eq}$ is roughly 1250 K, an order of magnitude above that of Jupiter and independent of $R_p$ and $M_p$. We assumed that the luminosity of 51 Peg A is 60% higher than that of our Sun (Mayor & Queloz 1995) and that absorbed heat is quickly distributed to the night side to be radiated ($f = 1$). The latter assumption is fully justified for a gas giant or for any planet with a thick atmosphere, due to rapid zonal and meridional circulation patterns, but may be problematic for a bare "rock." If 51 Peg B were a giant terrestrial planet without an atmosphere, its temperature at the sub-stellar point could be as high as 1500 K (Table 1b), above the melting point of many rocks. Needless to say, the planet's $L_{eq}$ is unaffected by tidal locking, though the phase dependence of its brightness is.

Figure 1 and Tables 1a & 1b show the pronounced and ***diagnostic*** variation of radius with composition. Planets composed of materials with low electron fraction per baryon and high Z are significantly more compact (Zapolsky & Salpeter 1969). A giant terrestrial planet would be three times smaller than a gas giant of the same mass, and its corresponding luminosity would be an order of magnitude lower. The latter depends upon the albedos assumed, but only weakly for albedos below 0.4. If photometry can be performed on 51 Peg B, a measurement of its bolometric luminosity would immediately distinguish the different models.

### 3. The H-R Diagram for Giant Planets

Figure 2 is a theoretical Hertzprung-Russell diagram that portrays the major results of this study. Depicted are $L-T_{eff}$ tracks[†] for the evolution of a 1 $M_J$ gas giant and $L$ for a 1 $M_J$ "olivine" planet (open triangles), all at a variety of orbital distances (indicated by the arrows). Also shown are the Hayashi (1961) track (boundary of the dark shaded region), the Hayashi exclusion zone (the dark shaded region itself), the Roche exclusion zone (the lightly shaded region), and the classical Jeans evaporation limit (dash-dotted line). The shape of the Hayashi exclusion zone and the evolutionary ages depend slightly upon the atmospheric model employed. Figures such as Figure 2 can be rendered for any specific planetary mass, albedo, and primary, but we focus here on $M_p = 1$ $M_J$ , $A = 0.35$, and 51 Peg A. The dashed lines on Figure 2 are lines of constant radius. The numbers on the tracks are the common logarithms of the ages in years.

The evolution of a fully convective planet can be separated into two phases: (1) a rapid contraction phase, with large internal luminosity (converted from potential gravitational energy) and increasing effective temperature (the Hayashi boundary from the top right to the top middle of Figure 2), and (2) a slow cooling phase during which both the internal luminosity and the effective temperature decrease (the Hayashi boundary from the top center to the bottom right of Figure 2). The transition between these two phases occurs at $R_p$'s around $4\,R_J$, regardless of the mass of the planet. The planet's internal luminosity tends to zero and its effective temperature tends to $T_{eq}$. The present Jupiter is depicted by a diamond in the lower right-hand corner of Figure 2. Its evolutionary track closely follows the convective Hayashi track.

[†]Here, again, luminosities do not include the reflected component.

For a given mass and composition, every fully convective model lies on the same curve in the H-R diagram. No model can exist to the right of this curve (at lower $T_{eff}$). This region, the dark-shaded zone to the right in Figure 2, is the Hayashi forbidden zone. (As a corollary, any planet that lies to the left of the curve is partially radiative/conductive.) This implies that fully convective models cannot exist for large effective temperatures ($T_{eff} > 1400$ K for $M_p = 1\,M_J$). However, as $T_{eff}$ approaches $T_{eq}$, its internal luminosity drops until a radiative zone appears in the outer region and grows. This allows the planet to cool and shrink beyond the limit set by fully convective calculations. This behavior is illustrated in Figure 2 by the curling of the lines off of the Hayashi track. At $0.05\,$A.U., a $1\,M_J$ planet follows the fully-convective track for less than $10^7$ years. It then has a radius of about $2.5\,R_J$. At that point, a radiative outer region appears and the planet slowly contracts at a nearly constant effective temperature. After 8 billion years of evolution, its radius is only $1.2\,R_J$ and its luminosity is about $3.5 \times 10^{-5}\,L_\odot$ (more than $1.5 \times 10^4$ times the present luminosity of Jupiter and only a factor of two below that at the edge of the main sequence). The radiative region encompasses the outer 0.03% in mass, and 3.5% in radius. The temperature is about $3080\,$K at $10\,$bar, and around $3.7 \times 10^4\,$K at the center of the planet.

The quasi-static evolution of partially radiative planets is possible even for tiny star-planet separations. Such models are not unstable. At small orbital separations, the evolution is substantially slowed down by stellar heating. The almost vertical evolution tracks seen in Figure 2 for orbital distances smaller than $0.04\,$A.U. are a consequence of the fact that the internal luminosity of the planet is constrained to be small. Otherwise, it would be fully convective, which is not possible at these effective temperatures. Similarly, the planet must contract slowly to keep the internal luminosity small. However, note that the lines are illustrative evolutionary tracks, started for specificity at high $L$'s and $T_{\text{eff}} = T_{\text{eq}}$.

The Roche-excluded region is bounded by a line of nearly constant $L$, whose value is proportional to $M_{\text{p}}^{2/3}$. This small, but significant, region constrains models for 51 Peg B's formation. If 51 Peg B were formed beyond an A.U. and moved inward on a timescale greater than $\sim 10^8$ years, it would closely follow the $R_{\text{p}} = \text{R}_{\text{J}}$ trajectory to its equilibrium position on Figure 2.

## 4. Thermal and Non-thermal Evaporation of a Gas Giant

If 51 Peg B is a gas giant, is it stable to evaporation and, if so, what is its current evaporation rate? We consider two potential loss mechanisms: (1) classical Jeans evaporation and (2) the non-thermal production of hot hydrogen atoms and ions by absorption of ultraviolet radiation from 51 Peg A.

The classical Jeans escape flux is proportional to $e^{-\lambda}(\lambda + 1)$, where $\lambda = GM_{\text{p}}m_{\text{H}}/kTR_{\text{p}}$ (Chamberlain and Hunten 1987). Here $m_{\text{H}}$ is the mass of the hydrogen atom or molecule, $k$ is Boltzmann's constant, and $T$ is the temperature of the planet at the escape level. For atomic hydrogen, if $T = 1300$ K, $R_{\text{p}} = 3$ R$_{\text{J}}$ , and $M_{\text{p}} = 0.5$ M$_{\text{J}}$ , $\lambda$ is close to 30 and Jeans escape might be important. The dash-dotted line on Figure 2 is the $\lambda = 30$ line. However, this combination of parameters is unlikely for 51 Peg B (see Figures 1 and 2). Our hydrogen-helium giant models at the age of 51 Peg A have radii closer to 1.2–1.3 R$_{\text{J}}$ , and actual $\lambda$'s between 65 and 280. Hence, our model planets in the Mayor & Queloz mass range are much too compact for classical Jeans escape of any ion or atom to be significant.

The production and escape of hot ions ($H^+$ and $H_2^+$) and hot atomic hydrogen by stellar ultraviolet radiation is much more likely, since these fragments obtain a residual, non-thermal, kinetic energy during their production. This leaves them in the fast tail of the Jeans escape function. Using the estimate of Atreya (1986) of $3 \times 10^9 \ cm^{-2}s^{-1}$ for the total $H^+$, $H_2^+$, and H flux from Jupiter, and assuming that the EUV flux from 51 Peg A is the same as the Sun's, we find that a gas giant at 0.05 A.U. with a mass of 1 $M_J$ would lose $10^{34}$ H's $s^{-1}$, or $10^{-16}$ $M_\odot \ yr^{-1}$. Only $\sim 0.5\%$ of the mass of a 1.0 $M_J$ gas giant at the position of 51 Peg B would be lost due to EUV radiation over the main sequence lifetime of 51 Peg A. Since the mechanical luminosity of the solar wind is similar to the Sun's total EUV luminosity, extrapolating the Sun's wind power to 51 Peg A implies that wind ablation of 51 Peg B may be no more important. (Note that a planet composed of a higher-Z material would be much less prone to evaporation or stripping.) Interestingly, the EUV evaporation rate of 51 Peg B may exceed 1% of the mass loss rate from 51 Peg A itself.

While these numbers suggest that a Jupiter-type planet at 0.05 A.U. is stable, they are only about two orders of magnitude away from erosion of the entire planet. Close observation may reveal rather dramatic phenomena associated with the escape from 51 Peg B of the dissociation and ionization products of $H_2$.

## 5. Model Diagnostics

We have shown how luminosity and radius are the primary discriminants between gas-giant and giant-terrestrial planet models for 51 Peg B. However, it may someday be possible to identify spectral signatures which can directly characterize the composition and/or origin of the object.

A primarily silicate, but Jovian-mass, planet is an unusual object which we cannot rule out. As Figures 1 & 2 and Tables 1a & 1b demonstrate, its luminosity would be one tenth that of a gas giant of the same mass. Its spectroscopic signature would be a strong silicate absorption band in the 10-micron wavelength region. A massive water vapor atmosphere would long ago have been photodissociated into hydrogen and oxygen, unless the abundance of water were a significant fraction (1%) of the mass of the planet. We consider this unlikely. The presence of molten sulfur compounds on the surface cannot be ruled out, and might even obscure completely the silicate spectral signature.

A predominantly hydrogen-helium planet will not appear like Jupiter, even if the composition is similar. Ammonia clouds in Jupiter's atmosphere play a significant role in determining the scattering properties of the atmosphere and help to shape and define the 5-micron spectral window region. Such clouds will be absent in 51 Peg B, as will water clouds. At an effective temperature of roughly 1250 K, the primary cloud-forming materials near the surface are magnesium silicates and other silicate compounds. However, our models suggest that these clouds will be below the unity optical depth level. Because of this, the scattering optical depth of the atmosphere is expected to be small, and absorption features relatively deep and well-defined. Collision-induced molecular hydrogen opacity will be an important source of absorption in the near-infrared and infrared and may provide detectable features. Importantly, water and carbon monoxide absorption features should be present (Lunine *et al.* 1986). In contrast to Jupiter, methane is expected to be absent spectroscopically, because at high temperatures carbon monoxide is the thermodynamically-preferred carbon-bearing molecule.

We have demonstrated in this paper that gas giants can be stable, even for very small orbital distances, and have explored the structural and thermal consequences of various models of 51 Peg B. Photometry and spectrophotometry, using very advanced interferometric and adaptive-optics techniques, may well be the key to distinguishing the different theories for the origin and nature of 51 Peg B (Angel 1994; Kulkarni 1992). However, whatever its true nature, 51 Peg B has opened a new chapter in planetary studies.

## 6. Acknowledgments

The authors would like to thank Willy Benz, Peter Hauschildt, Jim Liebert, Doug Lin, Geoff Marcy, Jay Melosh, and Andy Nelson for stimulating conversations, numbers in advance of publication, or both. Gratitude is extended to the N.S.F. and to N.A.S.A. for support under grants AST93-18970 and NAG 5-2817, respectively. T.G. acknowledges support from the European Space Agency and D.S. acknowledges NASA grant HF-1051.01-93A from the Space Telescope Science Institute, which is operated by the Association of Universities for Research in Astronomy, Inc., under NASA contract NAS5-26555.

Table 1a.   Radiative/convective H/He planet orbiting at 0.05 A.U. from 51 Peg A

| $M_{\rm p}/M_{\rm J}$ | $R_{\rm p}/R_{\rm J}$ | $T_{\rm eff}$ (K) | $\log(L/L_\odot)$ |
|---|---|---|---|
| 0.5 | 1.39 | 1241 | −4.39 |
| 1.0 | 1.25 | 1241 | −4.48 |
| 1.5 | 1.22 | 1241 | −4.50 |
| 2.0 | 1.21 | 1241 | −4.51 |

Table 1b.   Olivine planet orbiting at 0.05 A.U. from 51 Peg A. ("syn" is short for "synchronous" and implies that the peak values for the star-facing hemisphere are given.)

| $M_{\rm p}/M_{\rm J}$ | $R_{\rm p}/R_{\rm J}$ | $T_{\rm eff}$ (K); non-syn | $T_{\rm eff}$ (K); syn | $\log(L/L_\odot)$ |
|---|---|---|---|---|
| 0.5 | 0.31 | 1241 | 1476 | −5.69 |
| 1.0 | 0.34 | 1241 | 1476 | −5.61 |
| 1.5 | 0.35 | 1241 | 1476 | −5.59 |
| 2.0 | 0.35 | 1241 | 1476 | −5.59 |

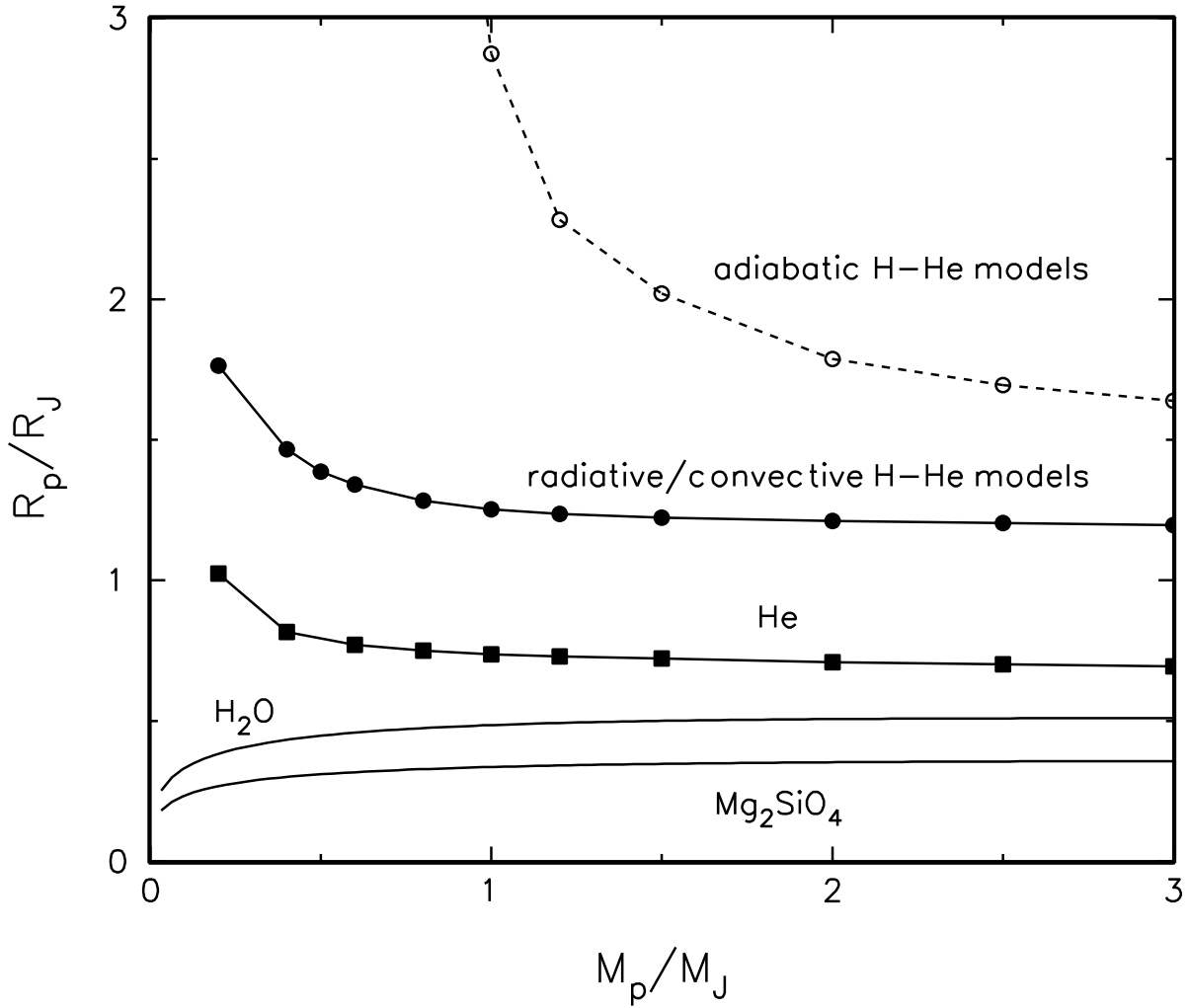

Fig. 1.— Radius ($R_p$) versus mass ($M_p$) for (top to bottom): fully adiabatic gas giants with surface temperature determined by radiative equilibrium with 51 Peg A; gas giants with radiative regions near the surface at the age of 51 Peg A (realistic gas-giant model); pure-helium giants with radiative/convective structure at the same age; pure $H_2O$ models at zero temperature; pure olivine ($Mg_2SiO_4$) models at zero temperature. The structures of the $H_2O$ and olivine planets were determined using the ANEOS equation of state (Thompson 1990). (The temperature corrections to the radii of these compact planet types are small.)

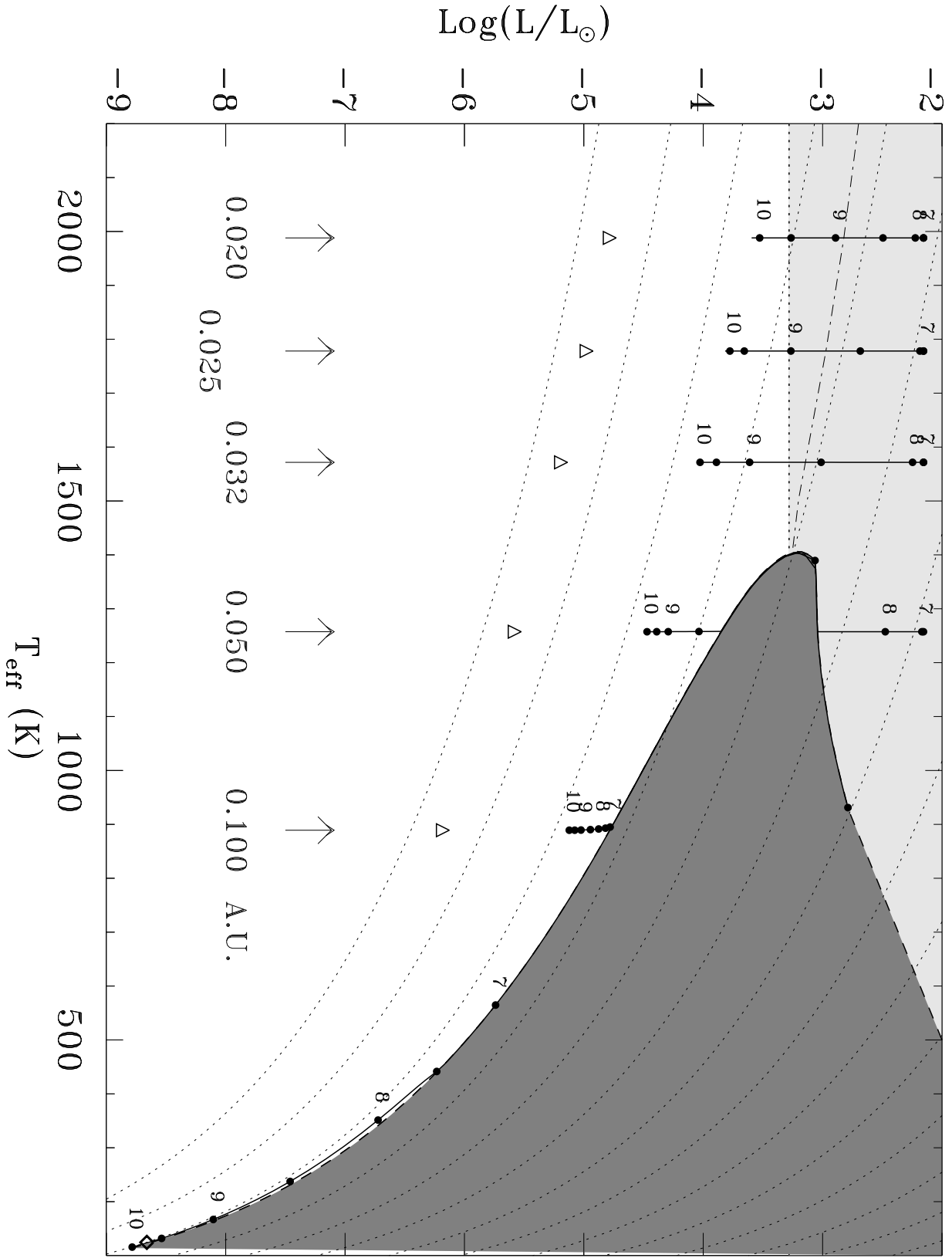

Fig. 2.— Hertzsprung-Russell diagram for $1 M_J$ planets orbiting at 0.02, 0.025, 0.032, 0.05, and 0.1 A.U. from a star with the properties of 51 Peg A, assuming a Bond albedo of 0.35. These can be scaled for different albedos. A model with $D$=0.1 A.U., $A$=0.35 is identical to one with $D$=0.05, $A$=0.84 (see Eq. 2). Arrows indicate the corresponding equilibrium effective temperature. A Jupiter model is also shown, the diamond in the bottom right-hand corner corresponding to the present-day effective temperature and luminosity of the planet. Evolutionary tracks for planets of solar composition are indicated by lines connecting dots which are equally spaced in log(time). The numbers 7, 8, 9, 10 are the common logarithms of the planet's age. Zero temperature models for $1 M_J$ planets made of olivine ($Mg_2SiO_4$) are indicated by triangles. The Hayashi forbidden region, which is enclosed by the fully convective model's evolutionary track is shown in dark grey (see text). Models in the light grey region assume an albebo of 0.35 and have radii above the Roche limit (and therefore are tidally disrupted by the star). The region where classical Jeans escape becomes significant is bounded by the dash-dotted line. Lines of constant radius are indicated by dotted curves. These correspond, from bottom to top, to radii (in units of $R_J$ ) in multiples of 2, starting at 1/4.